\documentclass[aps,prx,10pt,twocolumn,superscriptaddress]{revtex4}

\usepackage{graphicx}
\usepackage{amssymb, amsmath}
\usepackage{color}
\usepackage{ulem}
\usepackage{bm}
\usepackage{graphicx}
\usepackage{subfigure}
\usepackage{dcolumn}
\usepackage{mathtools}
\def\I{\uppercase\expandafter{\romannumeral 1}}
\def\II{\uppercase\expandafter{\romannumeral 2}}
\def\III{{\uppercase\expandafter{\romannumeral 3}}}
\def\IV{{\uppercase\expandafter{\romannumeral 4}}}
\def\V{{\uppercase\expandafter{\romannumeral 5}}}
\def\VI{{\uppercase\expandafter{\romannumeral 6}}}
\def\VII{{\uppercase\expandafter{\romannumeral 7}}}
\def\i{\lowercase\expandafter{\romannumeral 1}}
\def\ii{\lowercase\expandafter{\romannumeral 2}}
\def\iii{{\lowercase\expandafter{\romannumeral 3}}}
\def\iv{{\lowercase\expandafter{\romannumeral 4}}}
\def\v{{\lowercase\expandafter{\romannumeral 5}}}
\def\vi{{\lowercase\expandafter{\romannumeral 6}}}
\def\vii{{\lowercase\expandafter{\romannumeral 7}}}

\def\nn{\nonumber\\}

\def\k{\mathbf{k}}
\def\nn{\nonumber\\}

\def\k{\mathbf{k}}

\begin{document}

\title{Anomalous Hall effect, magneto-optical properties, and nonlinear optical properties of twisted graphene systems}

\author{Jianpeng Liu}
\affiliation{Department of Physics, Hong Kong University of Science and Technology, Kowloon, Hong Kong}

\author{Xi Dai}
\affiliation{Department of Physics, Hong Kong University of Science and Technology, Kowloon, Hong Kong}

\begin{abstract}
We study the anomalous Hall effect, magneto-optical properties, and nonlinear optical properties of twisted bilayer graphene (TBG) aligned with hexagonal boron nitride (hBN) substrate as well as twisted double bilayer graphene systems. We show that non-vanishing  valley polarizations in twisted graphene systems would give rise to anomalous Hall effect which can be tuned by in-plane magnetic fields. The valley polarized states are also associated with giant Faraday/Kerr rotations in the terahertz frequency regime. Moreover, both hBN-aligned TBG and TDBG exhibit colossal nonlinear optical responses by virtue of the inversion-symmetry breaking, the small bandwidth, and the small excitation gaps of the systems. Our calculations indicate that in both systems the nonlinear optical conductivities of the shift currents are on the order of $10^3\,\mu$A/V$^2$; and the second harmonic generation (SHG) susceptibilities are on the order of  $10^6\,$pm/V in the terahertz frequency regime.
Moreover, in TDBG with $AB\textrm{-}BA$ stacking, we find that a finite orbital magnetization would generate a new component $\sigma^{x}_{xx}
$ of the nonlinear  photoconductivity tensor; while in $AB$-$AB$ stacked TDBG with vertical electric fields, the valley polarization and orbital magnetization would make significant contributions to the $\sigma^{y}_{xx}$ component of the photoconductivity tensor. These nonlinear photo-conductivities are proportional to the orbital magnetizations of the systems, thus they are expected to exhibit hysteresis behavior in response to out-of-plane magnetic fields.

\end{abstract}

\maketitle

Twisted bilayer graphene (TBG)  has drawn significant attention recently due to the observations of the correlated insulating phases, anomalous Hall effect, and unconventional superconductivity \cite{cao-nature18-mott,sharpe-tbg-19,choi-tbg-stm,kerelsky-tbg-stm,marc-tbg-19,efetov-tbg-arxiv19,tbg-qah-young-arxiv19}.
At small twist angles, the low-energy states of  TBG are characterized by  two low-energy bands for each valley and spin degrees of freedom \cite{santos-tbg-prl07,macdonald-pnas11}. Around the ``magic angles", the bandwidths of the  low-energy bands become very small, and these nearly flat bands are believed to be responsible for most of the unconventional properties observed in TBG. 
Numerous theories have been proposed to understand the intriguing phenomena observed in TBG \cite{po-prx18, yuan-prb18, koshino-prx18, kang-prx18, xu-balents-prl18, zhang-tbg-prb19, song-tbg-18, yang-tbg18, po-tbg2, origin-magic-angle-prl19, pal-kindermann-arxiv18, angeli-tbg-prb18, jpliu-prb19, ll-tbg-lian,zhang-tbg-arxiv19, wolf-tbg-arxiv19,sboychakov-arxiv-18, isobe-prx18, xu-lee-prb18, huang-arxiv-18,liu-prl18,rademaker-prb18, venderbos-prb18, kang-tbg-prl19, xie-tbg-2018, jian-moire,tbg-coupled-wire, zaletel-tbg-2019, nematic-tbg-arxiv18, randeria-tbg-arxiv18, wu-chiral-tbg-prb19, roy-tbg-prb19, valley-jahn-teller-tbg, wu-xu-arxiv19,ashvin-nematic-arxiv19, hu-tbg-arxiv19,bernevig-tbg-superfluid-arxiv19}.

Recently unconventional superconducting  and correlated insulating behavior, as well as quantum anomalous Hall effect have been observed in twisted double bilayer graphene (TDBG) \cite{zhang-double-bilayer-arxiv19, kim-double-bilayer-arxiv19, cao-double-bilayer-arxiv19} and trilayer graphene with hexagonal boron nitride (hBN) substrate \cite{chen-trilayer-superconduct-arxiv19, chen-trilayer-hbn-arxiv19}, which has stimulated extensive theoretical interests \cite{zhang-trilayer-hbn-prb19, ashvin-double-bilayer-arxiv19, jpliu-tmg-arxiv19, koshino-tdbg-prb19,jung-arxiv19, yazyev-tdbg-arxiv19}. In particular, it has been proposed that topological flat bands generically exist in twisted double bilayer \cite{ashvin-double-bilayer-arxiv19, jpliu-tmg-arxiv19, koshino-tdbg-prb19} and twisted multilayer graphene systems \cite{jpliu-tmg-arxiv19}, and that the topological flat bands with non-vanishing valley Chern numbers are associated with large and valley-contrasting orbital magnetizatons, which may lead to an orbital ferromagnetic state once the valley symmetry is broken either spontaneously or due to external magnetic fields \cite{jpliu-tmg-arxiv19}. The orbital ferromagnetic states are also believed to exist in TBG aligned with hBN substrate \cite{sharpe-tbg-19,efetov-tbg-arxiv19,xie-tbg-2018}, in which (quantum) anomalous Hall effect has been observed at 3/4 filling  of the flat bands around the magic angle \cite{sharpe-tbg-19, tbg-qah-young-arxiv19}.

In this context, it is natural to ask how to probe such orbital ferromagnetic states in experiments. Certainly the most salient signature of the orbital magnetic state is the anomalous Hall effect (AHE). In typical magnetic materials, anomalous Hall effect results from the interplay between spin ferromagnetism and spin-orbit coupling (SOC) \cite{ahe-rmp10}: time-reversal ($\mathcal{T}$) symmetry  is first broken in the spin sector leading to spin magnetizations, then the $\mathcal{T}$ symmetry breaking is transmitted from the spin sector to the orbital sector via SOC. In orbital ferroamgnetic systems, however, $\mathcal{T}$ symmetry is directly broken in the orbital sector, and the AHE does not require any microscopic SOC. One thus expects that the AHE in an orbital ferromagnetic metal would be much more conspicuous than that in a spin ferromagnetic metal, and that the AHE will be quantized as nonzero integers if the orbital magnet is insulating. Such argument also applies to magneto-optical effects. As light is directly coupled with the orbital degrees of freedom of electrons,  the magneto-optical effects in an orbital ferromagnet should be much more pronounced than those in a spin ferromagnet.  Therefore, we expect that there would be significant magneto-optical responses in both hBN-aligned TBG and TDBG. On the other hand, inversion symmetry is broken in both systems, which allows for nonlinear optical effects \cite{kraut-prb79, sipe-prb00}. 
It is intriguing to ask whether the orbital magnetization and valley polarization in TBG and TDBG can be probed by nonlinear optical responses. Even without orbital magnetism, the nonlinear optical properties of the twisted graphene systems from structural inversion-symmetry breaking is still an open question, which deserves a comprehensive study.

In this paper, we systematically study the AHE, magneto-optical properties, and nonlinear optical properties of both hBN-aligned TBG and TDBG. We find that in additional to AHE, there are also giant magneto-optical Kerr and Faraday rotations in both systems by virtue of the valley-symmetry breaking and orbital magnetizations. Therefore, we propose that the Faraday and Kerr rotations may be a powerful tool to detect the presence of orbital magnetism in twisted graphene systems. Moreover, we also study the nonlinear optical responses such as the shift current and second harmonic generation (SHG) in hBN-aligned TBG and TDBG. We find that both systems exhibit colossal nonlinear optical responses by virtue  of the  small bandwidths and the small excitation gaps of the twisted graphene systems.  To be specific, our calculations indicate that in either system, the shift-current conductivity $\sigma^{c}_{ab}(0)$ is on the order of $10^{3}\,\mu$A/V$^2$,  and the SHG susceptibility $\chi^{c}_{ab}(2\omega)$ is on the order of $10^6$\,pm/V in the terahertz frequency regime. 
In TDBG with $AB$-$BA$ stacking, we propose that the non-vanishing valley polarization (orbital magnetization) would generate a new component of the nonlinear  photoconductivity $\sigma^{x}_{xx}$; while in $AB$-$AB$ stacked TDBG with vertical electric fields, we find that the orbital magnetization would make significant contributions to the $\sigma^{y}_{xx}$ component.  We predict that both of these two components of photoconductivities will exhibit remarkable hysteresis behavior in response to out-of-plane magnetic fields. Therefore, these nonlinear photo-conductivities that are generated by the orbital magnetizations may be considered as strong experimental evidence for the valley polarizations and orbital ferromagnetism in the TDBG systems.

\section{The TBG system aligned with hexagonal BN substrate}
We first study the AHE, magneto-optical properties, and nonlinear optical properties of the hBN-aligned TBG system.  We consider the situation that TBG is placed on top of a hBN substrate, and the hBN substrate is aligned with the bottom graphene layer. This is actually the device used in Ref.~\onlinecite{sharpe-tbg-19}, in which an anomalous Hall conductivity $\sim 2.4 e^2/h$ has been observed at 3/4 filling of the conduction flat band around the magic angle. The hBN substrate is believed to have two effects on the electronic structures of TBG. First, the alignment of the hBN substrate with the bottom graphene layer would impose a staggered sublattice potential on the bottom layer graphene and break the $C_{2z}$ symmetry, which  opens a gap at the Dirac points of the flat bands of the magic-angle TBG. Actually the two flat bands for the $K$ valley acquires nonzero Chern numbers $\pm 1$ ($\mp 1$ for the $K'$ valley) once a gap is opened up at the Dirac points. Second, the hBN substrate would generate a new moire pattern, which roughly has the same period as the one generated by the twist of the two graphene layers, but are orthogonal to each other \cite{moon-hbn-graphene-prb14}. However, the moir\'e potential generated by the hBN substrate is one order of magnitude weaker than that generated by the twist of the two graphene layers \cite{moon-hbn-graphene-prb14, jung-hbn-graphene-prb14}. Therefore, as a leading-order approximation, it is legitimate to neglect the moir\'e potential generated by the hBN substrate \cite{zhang-tbg-arxiv19, zaletel-tbg-2019}. 
 With such an approximation, the effective Hamiltonian for the hBN-aligned TBG system is simplified as
\begin{equation}
H^{\mu}_0=H_{TBG}^{\mu}+H_{mass}
\label{eq:h0}
\end{equation}
where $\mu\!=\!\pm 1$ is the valley index, and $H_{TBG}^{\mu}$ represents the continuum  Hamiltonian for valley $\mu$ as proposed by Bistrizer and MacDonald \cite{macdonald-pnas11}, 
and $H_{mass}$ is the ``Dirac mass" term at the bottom layer graphene generated by the hBN substrate, which is expressed as
\begin{equation}
H_{mass}=\begin{pmatrix}
\Delta\,\sigma_z & 0 \\
0 & 0 
\end{pmatrix}\;,
\label{eq:hmass}
\end{equation}
where $\Delta$ is the staggered sublattice potential exerted on the bottom graphene layer, which is fixed as $17\,$meV throughout this paper. The details of the continuum Hamiltonian of TBG $H_{TBG}^{\mu}$ is given in Supplementary Material.

\subsection{Electronic structures}
\begin{figure}
\includegraphics[width=3.5in]{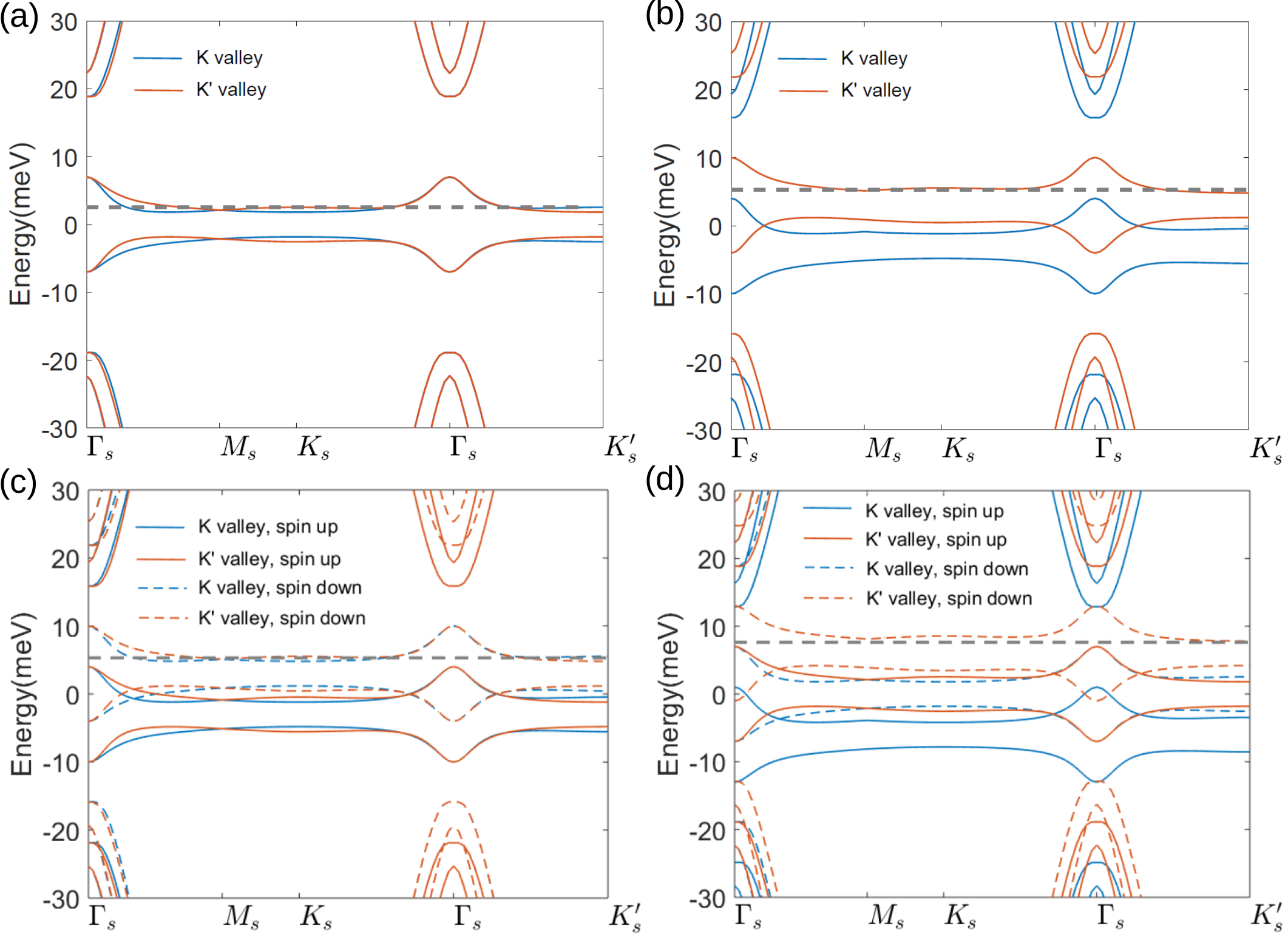}
\caption{Bandstructures of hBN-aligned TBG: (a) $E_v=0$, $E_v=0$, (b)  $E_v=3\,$meV and $E_s=0$, (c) $E_v=0$ and $E_s=3\,$meV, (d) $E_v=3\,$meV and $E_v=3\,$meV.}
\label{fig:band}
\end{figure}

In order to study the 
effects of breaking the valley and spin symmetries, we artificially apply valley and spin energy splittings 
\begin{equation}
H^{\mu s}=H^{\mu}_0+\mu E_v \tau_z + s E_s s_z\;,
\label{eq:hfull}
\end{equation}
where $E_v$ and $E_s$ are positive real numbers denote the valley and spin splittings, and $\mu\!=\!\pm 1$ and $s\!=\!\pm 1$ represent the valley and spin degrees of freedom respectively. $\tau_z$ and $s_z$ both denote the third Pauli matrix, and are defined in the valley and spin subspace respectively. 
The bandstructures at the first magic angle $\theta\!=\!1.05^{\circ}$ are shown in Fig.~\ref{fig:band}, where the solid blue lines, dashed blue lines, solid red lines, and dashed red lines denote the bandstructures of electrons with \{$\mu\!=\!-1$, $s\!=\!-1$\}, \{$\mu\!=\!-1$, $s\!=\!+1$\}, \{$\mu\!=\!+1$, $s\!=\!-1$\}, and \{$\mu\!=\!+1$, $s\!=\!+1$\} respectively. The thick dashed gray line denotes the chemical potential for some given valley and spin splittings, which is determined by the charge filling $+3/4$, i.e., filling 7 out of the 8 flat bands including the valley and spin degrees of freedom.  When $E_v\!=\!0$, $E_s\!=\!0$, the bandstructures are spin degenerate for each valley as shown in Fig.~\ref{fig:band}(a).  When $E_v\!=\!3\,$meV and $E_s\!=\!0$, the bandstructures are shown in Fig.~\ref{fig:band}(b). Clearly the two valleys have been splitted: the flat bands of the $K$ valley are completely filled, while  the conduction band of the $K'$ valley is half filled. Later we will show that the AHE, orbital magnetization, and magneto-optical effects would be maximal in such a situation.  In Fig.~\ref{fig:band}(c) we show the bandstructures with $E_v\!=\!0$ and $E_s\!=\!3\,$meV. We see that the spins have been splitted but the valley symmetry is still preserved. In this situation, both AHE and magneto-optical effects vanish since $\mathcal{T}$ symmetry is broken only in the spin sector, but still preserved in the orbital sector. In Fig.~\ref{fig:band}(d), we show the bandstructures with $E_v\!=\!3\,$meV and $E_s\!=\!3\,$meV at +3/4 filling. In this situation, both the spin and valley splittings are strong enough such that both the valley and spin polarizations $\xi_v$ and $\xi_s$ reach their maximal values with $\xi_v=\xi_s=1/7$   
\footnote{The valley and spin polarizations are defined as: $\xi_v=(\sum_{s=\pm 1}\rho_{\mu=-1,s}-\sum_{s=\pm 1}\rho_{\mu=+1,s})/\rho_t$, and $\xi_s=(\sum_{\mu=\pm 1}\rho_{\mu,s=-1}-\sum_{\mu=\pm 1}\rho_{\mu,s=+1})/\rho_t$, where $\rho_{\mu,s}$ is the charge density for valley $\mu$ and spin $s$, and $\rho_t=\sum_{\mu, s=\pm 1}\rho_{\mu, s}$ is the total charge density.}
. 
Then the system enters a quantum anomalous Hall (QAH) insulating phase with Chern number $-1$. Such a phase has been predicted as the ground state at +3/4 filling of magic-angle TBG based on Hartree-Fock calculations including electrons' Coulomb interactions \cite{xie-tbg-2018,zhang-tbg-arxiv19}.  

It is  important to note that as a result of the staggered sublattice potential from the hBN substrate, a gap $\sim\!4\,$meV  has opened up at the Dirac points $K_s$ and $K_s'$ as clearly shown in Fig.~\ref{fig:band}. Such a  $C_{2z}$ symmetry breaking and the gap opening at the Dirac points are essential in achieving AHE, magneto-optical effect and nonlinear optic effects in the TBG system. If $C_{2z}$ symmetry is preserved, the Hamiltonian for each valley and each spin would have $C_{2z}\mathcal{T}$ symmetry ($\mathcal{T}$ denotes time-reversal), which would enforce the Berry curvature to be zero at every $\k$ point. As a result, both the AHE and magneto-optical effects would be forbidden. The $C_{2z}$ operation also connects the two valleys. If $C_{2z}$ symmetry is preserved, the nonlinear optical response is also prohibited as the contributions from the two valleys would exactly cancel each other.

\subsection{Anomalous Hall effect}
\begin{figure}
\includegraphics[width=3.5in]{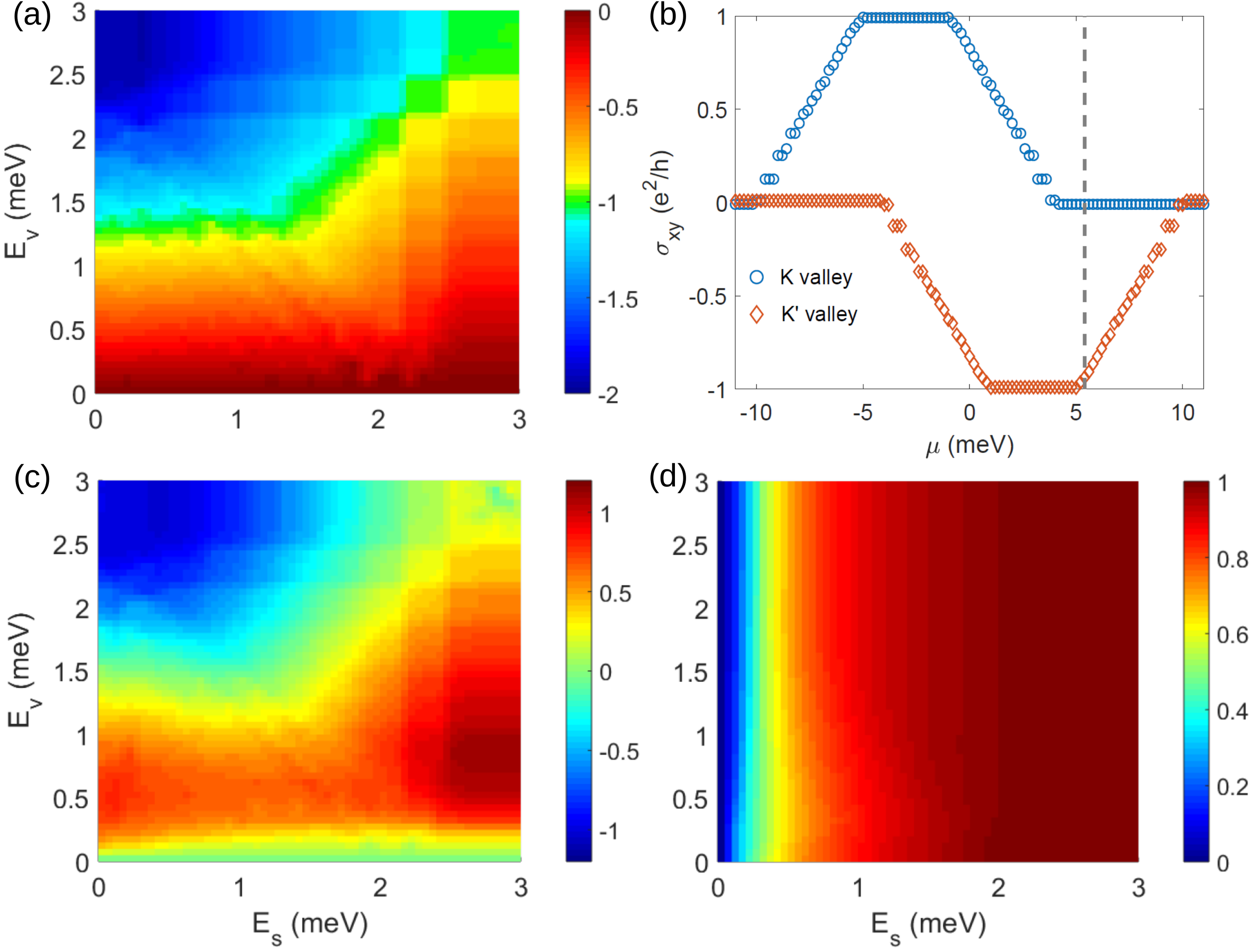}
\caption{(a) $\sigma_{xy}$ (in units of $e^2/h$) of 3/4-filled TBG aligned with hBN substrate at the magic angle.  (b) The frequency dependence of $\sigma_{xy}$ with $E_v\!=\!3\,$meV, $E_s\!=\!0$. (c) The orbital magnetization, and (d) the spin magnetization of 3/4-filled hBN-aligned TBG at the magic angle, in units of $\mu_B$ per moir\'e unit cell. The vertical and horizontal axes in (a), (c) and (d) are the valley and spin splittings respectively, in units of meV.}
\label{fig:ahc}
\end{figure}

We take the valley and spin splittings ($E_v$ and $E_s$ in Eq.~(\ref{eq:hfull})) as two free parameters which are varied from 0 to $3\,$meV. Then we study the dependence of AHE, spin/orbital magnetizations, and magneto-optical effects on the valley and spin splittings at +3/4 filling. In Fig.~\ref{fig:ahc}(a) we first show the dependence of the anomalous Hall conductivity (in units of $e^2/h$) on $E_v$ and $E_s$. Clearly the anomalous Hall conductivity (AHC) is non-vanishing only if $E_v\!>\!0$, otherwise $\mathcal{T}$ symmetry is always preserved in the orbital sector and the contributions from the two valleys always cancel each other. One may notice in Fig.~\ref{fig:ahc}(a) that there is a small region in the upper right corner in which $\sigma_{xy}$ is quantized as $-e^2/h$. This is the region in which both valley and spin polarizations reach their maximal values $\xi_v=\xi_s=1/7$, and the system enters a QAH phase with quantized Hall plateau. We note that $E_v\!\approx\!E_s\!\approx\!2.5\,$meV would be strong enough to approach the QAH phase. 

It is also interesting to note that when $E_v\!\sim\!3\,$meV and $E_s\!\sim\!0$\,meV, i.e., in the upper left corner of Fig.~\ref{fig:ahc}(a), the AHC is maximal with $\sigma_{xy}\!\sim\!-1.9\,e^2/h$, and the system is metallic as shown by the bandstructures in Fig.~\ref{fig:ahc}(b).  In other words, fixing $E_v\!\sim\!3$\,meV, $\sigma_{xy}$ would decrease from $-1.9\,e^2/h$ to $-e^2/h$ as $E_s$ increases from 0 to $3\,$meV, and the system would go through a transition from a metal to a QAH insulator. In Ref.~\onlinecite{sharpe-tbg-19}, the magnitude of the measured AHC $\sim 2.4\,e^2/h$ around +3/4 filling, and the system is still metallic. It may suggest that the valley splitting dominates over the spin splitting in their sample, such that the system still stays in a metiallic phase as shown Fig.~\ref{fig:band}(b), with  $\vert\sigma_{xy}\vert$ greater than $e^2/h$. According to our calculations, increasing the spin splitting may drive the system from the metallic phase to the spin and valley polarized QAH phase for the sample used in Ref.~\onlinecite{sharpe-tbg-19}. We expect that such a phase transition may be assisted by applying in plane magnetic field which only couples to the spin magnetization.

 In Fig.~\ref{fig:ahc}(b) we plot the dependence of $\sigma_{xy}$ on the chemical potential with $E_v\!=\!3\,$meV and $E_s\!=\!0\,$meV. The vertical gray dashed line marks the actual chemical potential at 3/4 filling. We see that as a result of the valley splitting, the bands from the $K$ valley are completely filled, and do not contribute to AHC. However, the conduction band of the $K'$ valley is only half filled. By virtue of the giant density of states near the conduction band minimum, the chemical potential is just slightly above the conduction band minimum, thus the AHC contributed by the $K'$ valley for each spin species is still close to the quantized value $-e^2/h$. This explains the large calculated AHC $\sigma_{xy}\!\sim\!-1.9\,e^2/h$ when $E_v\!\approx\!3\,$meV and $E_s\!\approx\!0$\,meV shown in Fig.~\ref{fig:ahc}(a). 
 
In Fig.~\ref{fig:ahc}(c)-(d) we plot the orbital and spin magnetizations (in units of $\mu_{B}$ per moir\'e primitive cell) in the parameter space spanned by $E_v$ and $E_s$. The orbital magnetization can be as large as $\sim -1\,\mu_{B}$ when $E_v\!\sim\!3\,$meV and $E_s\!\sim\!0\,$meV, and gradually decreases to $\sim 0.1\,\mu_B$ as $E_s$ increases to $\sim\!3\,$meV.  On the other hand, in the QAH phase with $E_v\!\approx\!E_s\!\sim\!3\,$meV, the spin magnetization is as large as $1\,\mu_B$ per moir\'e cell (see Fig.~\ref{fig:ahc}(d)), indicating that the magnetization in the QAH phase would be dominated by the spin component. However, the orbital magnetic order is extremely anisotropic, which breaks a discrete $\mathbb{Z}_2$ symmetry; while the spin magnetic order is isotropic due to the absence of atomic SOC, which breaks continuous spin rotational symmetry. Therefore, despite being small in magnitude, the orbital magnetization is expected to be much more robust to thermal fluctuations according to Mermin-Wagner theorem.

\subsection{Magneto-optical properties}

\begin{figure}
\includegraphics[width=3.5in]{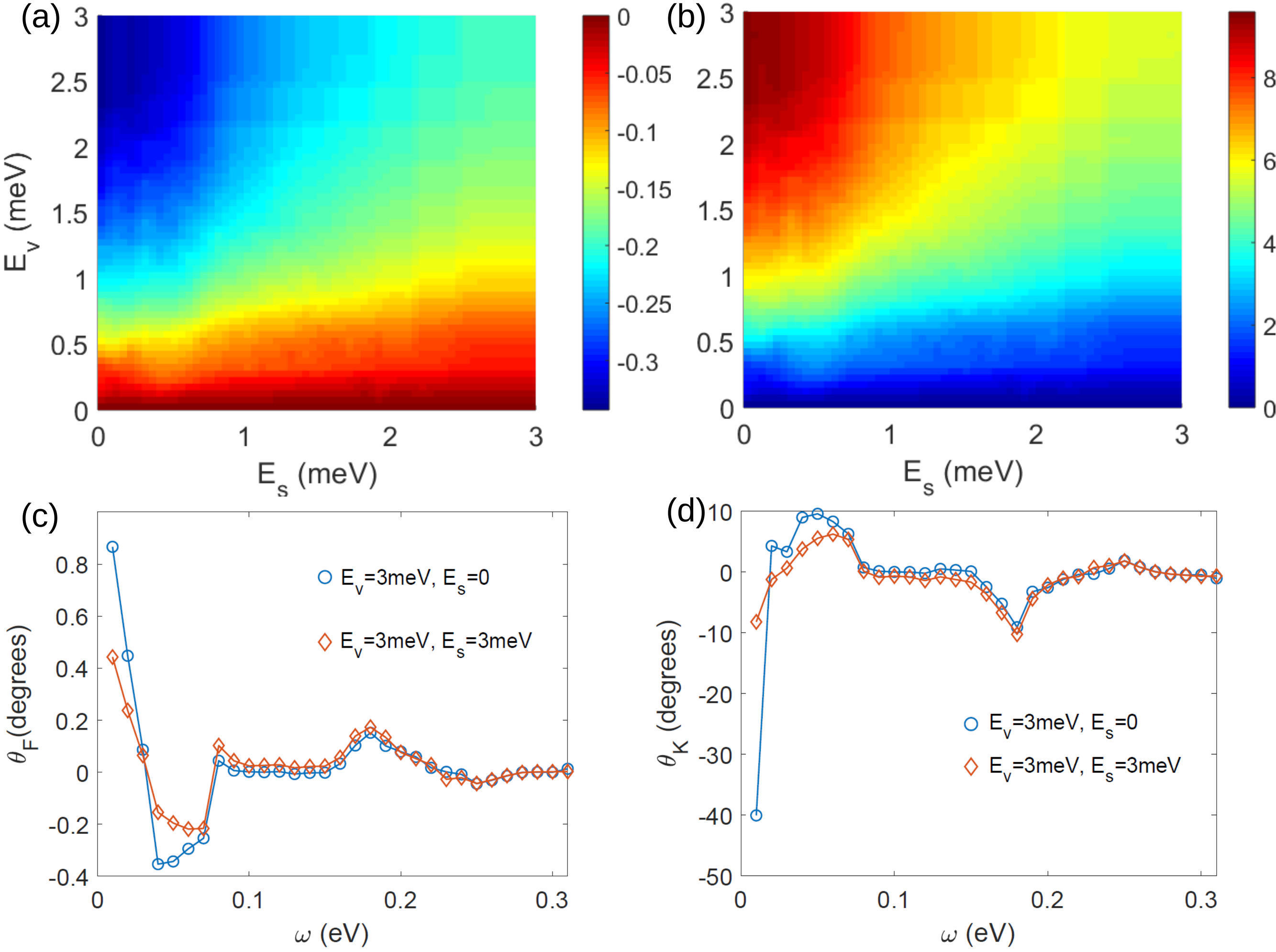}
\caption{(a)The Faraday angle $\theta_F$, and (b) the Kerr angle $\theta_K$, for hBN-aligned TBG at the magic angle at 3/4 filling. The vertical and horizontal axes are the valley and spin splittings respectively. For the same system, the frequency dependence of the Faraday angle (c), and the Kerr angle (d). The blue circles and red diamonds represent the situations with $\{E_v\!=\!3\,$meV, $E_s\!=\!0\}$, and $\{E_v\!=\!3\,$meV, $E_s\!=\!3\,$meV\}  respectively.}
\label{fig:kerr}
\end{figure}

In Fig.~\ref{fig:kerr}(a)-(b) we plot the Faraday and Kerr rotations (denoted by $\theta_F$ and $\theta_K$) as $E_v$ and $E_s$ increases from 0 to $3\,$meV (see supp. mat. for details). We consider the case that the incident light is normal to the 2D plane, and we first fix the frequency $\hbar\omega\!=\!0.05\,$eV. As clealry shown in the figures, both $\theta_F$ and $\theta_K$ vanish when $E_v=\!0\!$, and their magnitudes increase with the increase of the valley splittings. When $E_v\!\sim\!3$\,meV, $\theta_F\!\sim\!-0.4^{\circ}$ and $\theta_K$ is as large as $9^{\circ}$. In the QAH phase (upper right corner), $\theta_F\!\approx\!-0.2^{\circ}$ and $\theta_K\!\approx\!5.6^{\circ}$ with the incident light frequency $\hbar\omega\!=\!0.05\,$eV.  We note that  the calculated Kerr rotation in hBN-aligned TBG is at least an order of magnitude greater than those observed in typical spin ferromagnetic materials \cite{antonov-kerr-book}. For example, in $3d$ transition metal compounds such as Fe, Co, Ni, and MnPt$_3$, the maximal Kerr angles are typically  on the order of $0.1^{\circ}-1^{\circ}$ over the entire frequency regime \cite{antonov-kerr-book} ; in some magnetic multilayers and heterostructrures such as Co/Pd(Pt) multilayers \cite{antonov-kerr-book}, Fe/Au multilayers \cite{antonov-kerr-book}, and yttrium-iron garnet thin films \cite{yig-kerr-prl06},  the measured maximal $\theta_K$ is also on the order of $0.1^{\circ}-1^{\circ}$ \cite{antonov-kerr-book}. 
In double-layer CrGeTe$_3$, the Kerr rotation  $\theta_K\sim 0.0007^{\circ}$ at the light frequency $\hbar\omega\approx 0.35\,$eV \cite{crgete3-nature17}; while in single-layer CrI$_3$, the Kerr angle $\theta_K\sim 0.3^{\circ}$ at the frequency $\hbar\omega\approx 1.95\,$eV, which has been proposed to arise from excitonic effects \cite{wu-exciton-kerr}. We see that the calculated Kerr angle in hBN-aligned TBG in the terahertz regime is order of magnitude larger than those of any conventional  magnetic materials with small SOC. 

In typical spin ferromagnetic materials as mentioned above, the magneto-optical phenomena result from  the interplay between spin ferromagnetism and SOC: the SOC transimits the TR symmetry breaking from spin sector to orbital sector, and generates orbital magnetizations.  The orbital magnetization and magneto-optical effects would be  vanishingly small if SOC amplitude is negligible. However, in most magnetic materials, the effects of SOC are perturbative compared with the bandwidths. Therefore, the observed $\theta_K$ are typically very small ($\sim 0.1^{\circ}$) in  ferromagnetic transition-metal compounds. On the other hands, in twisted grapehen systems, the Faraday and Kerr rotations directly result from the orbital ferromagnetism as indicated by the significant orbital magnetization shown in Fig.~\ref{fig:ahc}(c). The spin and orbital magnetization coexist in the hBN-aligned TBG system and are interwined with each other, but there is no microscopic SOC at the single-particle level. Thus it is expected that the Kerr and Faraday rotations in hBN-aligned TBG (with valley symmetry breaking) would exhibit similar behavior as those in Landau levels \cite{quantum-faraday-kerr-graphene, morimoto-prl09,optical-hall-ll-prl10,volkov-faraday-ll} and in quantum anomalous Hall (QAH) insulators \cite{tse-prb11, tokura-nc16}.
However, in hBN-aligned TBG, the difference is that here we do not need external magnetic fields nor any SOC to generate the quantized (anomalous) Hall conductivity; instead, the valley/spin symmetry is expected to be broken spontaneously due to Coulomb interactions.

%

In Fig.~\ref{fig:kerr}(c)-(d) we plot the frequency dependence of $\theta_F$ and $\theta_K$  for \{$E_v=3\,$meV, $E_s=0$\} (blue circles) and \{$E_v=3\,$meV, $E_s=3$\,meV\} (red diamonds).
Both $\theta_F$ and $\theta_K$ increase dramatically as $\omega$ decreases. In particular, in the QAH phase ($E_v\!=\!3\,$meV, $E_s\!=\!3$\,meV) $\theta_F\!=\!0.44^{\circ}$ and $\theta_K\!=\!-8.2^{\circ}$ at $\hbar\omega\!=\!0.01\,$eV; while when $E_v\!=\!3\,$meV and $E_s\!=\!0$, $\theta_F\!=\!0.87^{\circ}$, and $\theta_K\!=\!-40.04^{\circ}$ for $\hbar\omega\!=\!0.01\,$eV. Actually for an QAH insulator with Chern number $C$, in the limit $\omega\to 0$, $\theta_F$ should be quantized as integer multiples of the fine-structure constant $C\alpha\approx C/137\,$rad$\approx C\times0.42^{\circ}$   \cite{qi-prb08, tse-prb11}, which is consistent with the results shown in Fig.~\ref{fig:kerr}(a). In the QAH phae, the Kerr angle is predicted to be quantized as  $\pm\pi/2$ in the low-frequency limit \cite{tse-prb11}. Indeed our results at lower frequencies  ($\hbar\omega < 0.01\,$eV) indicate that $\theta_K$ tend to approach $90^{\circ}$ as $\omega$ is approaching 0
\footnote{In our calculations we have chosen an artificial quasi-particle life time $\tau=\hbar/\delta$, with $\delta=3\,$meV, so that the results are reliable only for $\hbar\omega\gg \delta$. Therefore, the results at frequencies lower than $10\,$meV are not presented as they are no longer reliable when $\hbar\omega\!\sim\!\hbar/\tau$}.
. 
It worthwhile to note that both $\theta_F$ and $\theta_K$ would change  signs at $\hbar\omega\sim 0.035\,$eV and $\hbar\omega\sim0.075\,$eV. This is because the real part of the optical anomalous Hall conductivity $\sigma_{yx}(\omega)$ change sign at $\hbar\omega\approx 0.035\,$eV and $0.075\,$eV, leading to the sign change in the Faraday and Kerr rotations (see Supplemental Material for more details). The sign change of the Faraday and Kerr angles may be an interesting feature which can be easily verified experimentally.

\subsection{Nonlinear optical properties}
We continue to study the nonlinear optical properties of hBN-aligned TBG. In general, the photocurrent $j^{c}(\omega_3)$ is related to the time-dependent electric fields of the light via the second-order  photoconductivity: $j^{c}(\omega_3)=\sum_{ab}\sigma^{c}_{ab}(\omega_3)\,E_a(\omega_1)\,E_b(\omega_2)$, where $a, b, c=x,y$ denotes the spatial directions in Cartesian coordinates  
\footnote{We consider the case of normal incident light, so that the polarization of light is always within the 2D plane. Therefore, $a, b,c=x, y$, and cannot be along the $z$ direction.}
, and $\omega_1$ and $\omega_2$ are the frequencies of the two incident photons, and $\omega_1+\omega_2=\omega_3$ \cite{sipe-prb00}. For monochromatic light, the frequency of the incident photons is fixed as $\pm\omega$, thus there could be two distinct second-order optical processes with $\omega_3=0$ or $\omega_3=2\omega$, corresponding to the generation of the shift current and the second harmonic generation respectively.
Regardless of the microscopic mechanism, the nonlinear optical conductivity tensor $\sigma^{c}_{ab}(0)$ and $\sigma^{c}_{ab}(2\omega)$ have the same properties under symmetry operations, thus they have the same symmetry-allowed components. In particular, one may expand $\sigma^{c}_{ab}$ to the leading order of the orbital magnetization $M_z$,
\begin{equation}
\sigma^{c}_{ab}=\sigma^{c}_{ab,0}+\sigma^{c}_{ab,z}M_z\;.
\label{eq:shift-general}
\end{equation}
The only symmetry of hBN-aligned TBG is $C_{3z}$, which restricts $\sigma^{c}_{ab,0}$, $\sigma^{c}_{ab,z}$ and $\sigma^{c}_{ab,zz}$ to the following form
\begin{align}
&\sigma^{x}_{xx,0} = -\sigma^{x}_{yy,0}=-\sigma^{y}_{xy,0}=-\sigma^{y}_{yx,0}\;\nn
&\sigma^{y}_{xx,0} = \sigma^{x}_{xy,0}=\sigma^{x}_{yx,0}=-\sigma^{y}_{yy,0}\;\nn
&\sigma^{x}_{xx,z} = -\sigma^{x}_{yy,z}=-\sigma^{y}_{xy,z}=-\sigma^{y}_{yx,z}\;\nn
&\sigma^{y}_{xx,z} = \sigma^{x}_{xy,z}=\sigma^{x}_{yx,z}=-\sigma^{y}_{yy,z}\;,
\label{eq:shift-c3z}
\end{align}
It turns out that there are only two independent photoconductivities $\sigma^{x}_{xx}$ and $\sigma^{y}_{xx}$ for both shift-current and SHG second-order responses. Each of them include two components: one is independent of the orbital magnetization $M_z$, and the other is linear in $M_z$.
The component that is linear in $M_z$ may vary with perpendicular magnetic field, and show hysteresis behavior due to the hysteresis loop of the orbital magnetization. Such a hysteresis behavior will only show up with perpendicular magnetic field since the orbital magnetization is pointing along the $\pm z$ direction.

Microscopically, the shift-current photoconductivity $\sigma^{c}_{ab}(0)$ and the second-harmonic photoconductivity $\sigma^{c}_{ab}(2\omega)$ can be derived using second-order perturbation theory, which are expressed as \cite{zhang-photogalvanic-prb18}
\begin{widetext}
\begin{align}
&\sigma^{c}_{ab}(0)=\frac{e^3}{\omega^2}
\sum_{\Omega=\pm\omega}\sum_{lmn}\int \frac{d\k}{(2\pi)^d}\,\textrm{Re}\,[\,\phi_{ab}\,(f_l-f_n)\,\frac{v^{a}_{nl}\,v^{b}_{lm}\,v^{c}_{mn}}{(E_{n\k}-E_{m\k}-i\delta)\,(E_{n\k}-E_{l\k}+\hbar\Omega-i\delta)}\,]\;\nn
&\sigma^{c}_{ab}(2\omega)=\frac{e^3}{\omega^2}
\sum_{\Omega=\pm\omega}\sum_{lmn}\int \frac{d\k}{(2\pi)^d}\,\phi_{ab}\,(f_l-f_n)\,\frac{v^{a}_{nl}\,v^{b}_{lm}\,v^{c}_{mn}}{(E_{n\k}-E_{m\k}-2\hbar\Omega-i\delta)\,(E_{n\k}-E_{l\k}+\hbar\Omega-i\delta)}\;,
\label{eq:shift-shg}
\end{align}
\end{widetext}

and the susceptibility of SHG $\chi^{c}_{ab}(2\omega)\!=\!i\sigma^{c}_{ab}(2\omega)/(2\epsilon_0\,\omega)$.

\begin{figure}
\includegraphics[width=3.5in]{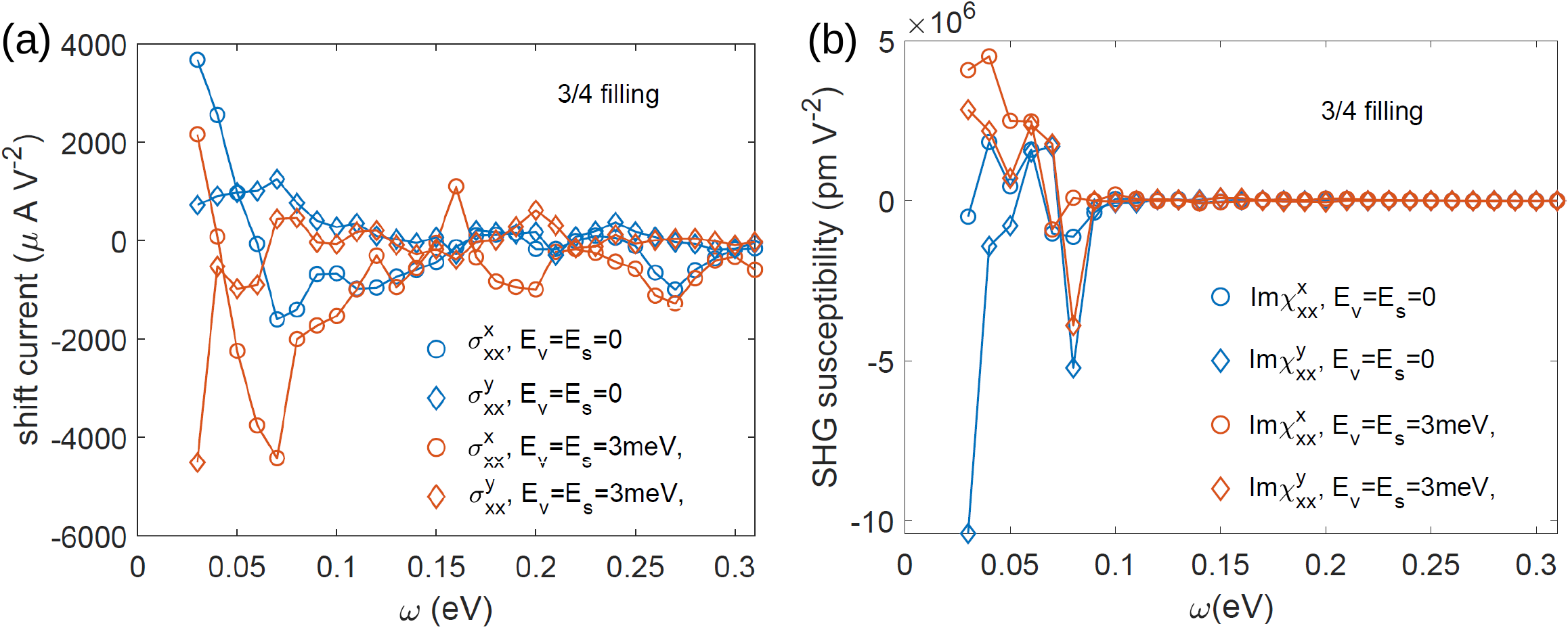}
\caption{(a) Shift-current photoconductivities $\sigma^{c}_{ab}(0)$, and (b) the imaginary parts of the SHG susceptibilities $\textrm{Im}\chi^{c}_{ab}(2\omega)$ for the 3/4-filled hBN-aligned TBG at the magic angle. The blue and red markers represent the cases of $E_v\!=\!=\!E_s\!=0$ and $E_v\!=\!=\!E_s\!=3\,$meV respectively. The circles and diamonds denote $\sigma^{x}_{xx}$ (or Im$\chi^{x}_{xx}$)  and $\sigma^{y}_{xx}$ (or Im$\chi^{y}_{xx}$) respectively.    }
\label{fig:shift-shg}
\end{figure}

In Fig.~\ref{fig:shift-shg}(a) we plot the frequency dependence of the shift-current photoconductivities at the +3/4 filling of hBN-aligned TBG, where the blue and red markers denote the situations with $E_v=E_s=0$ and $E_v=E_s=3\,$meV respectively, and the circles and diamonds represent $\sigma^{x}_{xx}(0)$ and $\sigma^{y}_{xx}(0)$ respectively. We note that the photoconductivities  are as large as $\sim\pm 4000\,\mu$A/V$^2$ at $\hbar\omega\!\lessapprox\!0.07$\,eV, which is unprecedentedly large. As the frequency increases, the photocondutvities can change sign and decrease to $\sim 10^{2}\,\mu$A/V$^2$ at relatively high frequencies $\hbar\omega\!\sim\!0.2-0.3\,$eV, which is comparable to the calculated shift-current conductivity of bilayer CrI$_3$ in the visible-light frequency regime \cite{cri3-shift-current-arxiv19}.  
In Fig.~\ref{fig:shift-shg}(b) we show the imaginary part of the SHG susceptibilities, where the blue and red markers denote situations with $E_v=E_s=0$ and $E_v=E_s=3\,$meV (the QAH phase), and the circles and diamonds represent $\chi^{x}_{xx}$ and $\chi^{y}_{xx}$ respectively.
At low frequencies $\hbar\omega\!\lessapprox\!0.1\,$ the SHG susceptibilities are extremely large, on the order of $10^6\,$pm/V, and they gradually decrease to $\sim\!10^3-10^4\,$pm/V at higher frequencies. Such colossal SHG susceptibilities 
are orders of magnitudes larger than those observed in other 2D materials with broken inversion symmetry such as monolayer MoS$_2$  \cite{mos2-shg-prb13, paula-mos2-shg-prb13}, monolayer WSe$_2$ \cite{xu-wse2-shg-13}, WS$_2$ \cite{ws2-shg-14}, and antiferromagnetic bilayer CrI$_3$ \cite{wu-cri3-shg-19}. In these 2D materials, the reported SHG susceptibilities are typically on the order of $10^3-10^5$\,pm/V in the visible-light frequency regime. 
The shift-current and SHG responses at other fillings are on the same order of magnitude as those of 3/4 filling, which we refer the readers to Supp. Mat. for details.

%

The  colossal shift-current and SHG responses shown in Fig.~\ref{fig:shift-shg} can be interpreted as follows. First, it is straightforward to see  from Eq.~(\ref{eq:shift-shg}) that the SHG and shift-current responses would be significantly enhanced when the one-photon or two-photon energy ($\omega$ or $2\omega$) is in resonance with some excited electronic states at some $\k$ points. Such a resonant enhancement would be much more pronounced if the energy bands are flat, as the flatness of the bands implies that the all electronic states at different $\k$ points would have the same resonant frequency, thus the resonance effect at different $\k$ points would be summed up. One could imagine having some sets of perfectly flat bands, such as Landau levels, but with broken inversion symmetry. Then if the frequency is in resonance with the Landau-level spacing, one would get enormous shift-current and SHG responses. Such giant nonlinear optical effects have never been observed in Landau levels because there is always inversion symmetry in Landau levels of free 2D electrons' gas or free Dirac fermions in graphene.
In hBN-aligned TBG, however, the low-energy states can be interpreted as pseudo Landau levels \cite{jpliu-prb19} with broken inversion symmetry due to the presence of the hBN substrate. These bands are roughly flat around the magic angle as shown in Fig.~\ref{fig:band}, which is expected to contribute to giant nonlinear optical responses once the incident photon frequency is somewhere in resonance with the electronic excitations. The electronic excitations in hBN-aligned TBG turn out to have small gaps, i.e., $\sim\!1\,$meV in the QAH phase at 3/4 filling, and $\sim 15\,$meV at the full filling (see Fig.~\ref{fig:band}), which indicates that the resonant frequencies in hBN-aligned TBG is very small. The small resonant frequencies would further amplify the giant nonlinear optical responses due to the $1/\omega^2$ dependence of the nonlinear photoconductivities (see Eq.~(\ref{eq:shift-shg})).

\section{Twisted double bilayer graphene} 


\subsection{Model Hamiltonian for TDBG}
The continuum Hamiltonian proposed for TBG \cite{macdonald-pnas11} can be further generalized to TDBG \cite{jung-arxiv19, ashvin-double-bilayer-arxiv19,koshino-tdbg-prb19, jpliu-tmg-arxiv19}:
\begin{equation}
H_{\lambda,\lambda'}^{\mu}=\begin{pmatrix}
h_1(\k) & h_{\lambda} & 0 & 0 \\
h_{\lambda}^{\dagger} & h_2(\k) & U_{\mu}(\mathbf{r}) & 0 \\
0 & U^{\dagger}_{\mu}(\mathbf{r}) & h_3(\k) & h_{\lambda'} \\
0 & 0 & h_{\lambda'}^{\dagger} & h_4(\k)
\end{pmatrix}\;,
\label{eq:Htdbg}
\end{equation}
where $h_l(\k)$ ($l=1,2$) denotes the low-energy effective Hamiltonian for monolayer graphene, i.e.,  $h_l(\k)=-v_F(\k-\mathbf{K}_l^{\mu})\cdot [\mu\sigma_x, \sigma_y] + (l-1)U_d/3$, where $\mathbf{K}_l^{\mu}$ are the $K$ or $K'$ point of the $l$th layer, $v_F$ is the bulk Fermi velocity, $U_d$ is the vertical electrostatic potential across the double bilayer, and $\mu=\pm 1$ is the valley index. $h_{\lambda}$ is the interlayer coupling of the $AB$ ($\lambda\!=\!+1$) or $BA$ ($\lambda\!=\!-1$) stacked bilayer graphene, and $U_{\mu}(\mathbf{r})$ is the moir\'e potential for valley $\mu$, which is  generated by the twist of the double bilayers. We refer the readers to Supp. Mat. for the explicit expressions of the interlayer coupling $h_{\lambda}$ and the moir\'e potential $U_{\mu}(\mathbf{r})$.
The effective Hamiltonian for TDBG with $AB$-$AB$ and $AB$-$BA$ stackings  would correspond to  $H_{+1,+1}^{\mu}$ and $H_{+1,-1}^{\mu}$ respectively. Eq.~(\ref{eq:Htdbg}) is the Hamiltonian for each valley and spin species. In order to study the effects of breaking the valley and spin symmetries, we apply artificial valley and spin splittings $H_{\lambda,\lambda'}^{\mu}$:
\begin{equation}
H^{\mu s}_{\lambda,\lambda;}=H^{\mu}_{\lambda,\lambda'}+\mu E_v \tau_z + s E_s s_z\;,
\label{eq:htdbg-eves}
\end{equation}
where $E_v$ and $E_s$ are positive real numbers denote the valley and spin splittings, and $\mu\!=\!\pm 1$ and $s\!=\!\pm 1$ represent the valley and spin degrees of freedom respectively. $\tau_z$ and $s_z$ both denote the third Pauli matrix, but defined in the valley and spin space respectively. In what follows we will study the dependence of AHC and magneto-optical effects on $E_v$ and $E_s$.

\subsection{Anomalous Hall effect and Magneto-optical properties}

\subsubsection{$AB$-$BA$ stacking}
\begin{figure}
\includegraphics[width=3.5in]{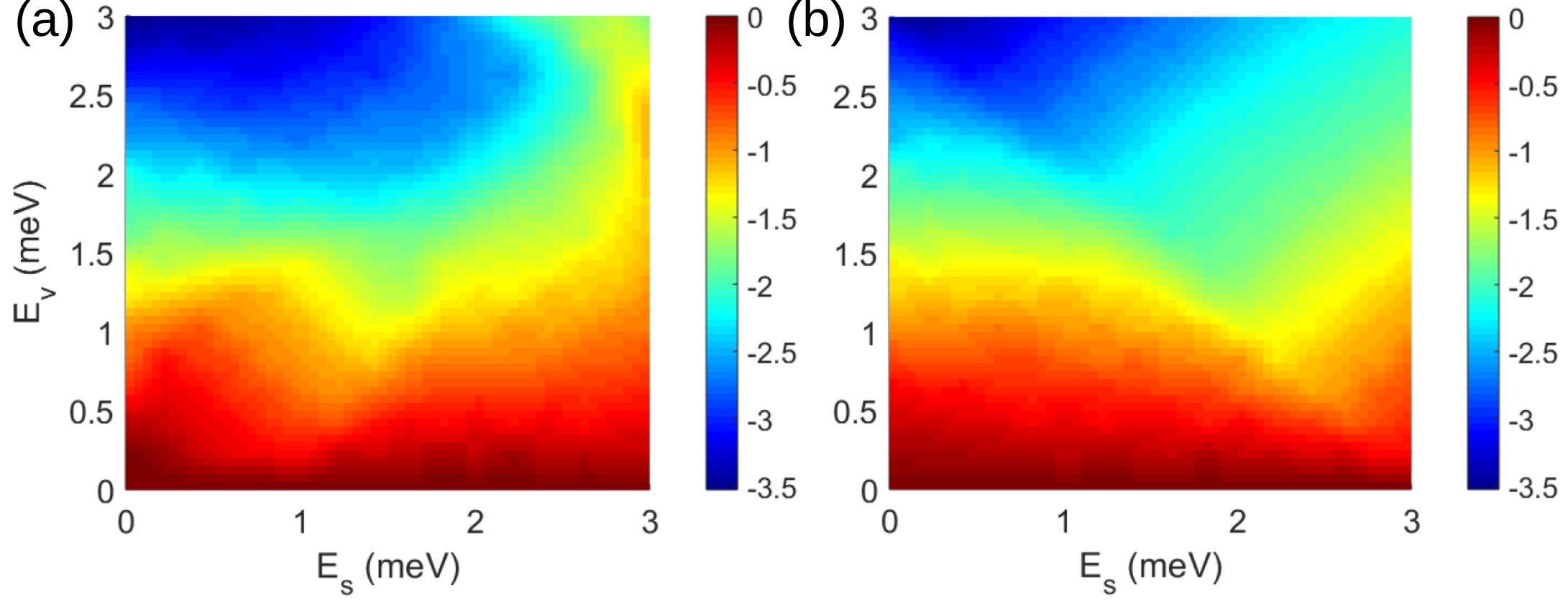}
\caption{$\sigma_{xy}$ for $AB$-$BA$ stacked TDBG, in units of $e^2/h$: (a) at 0 filling, and (b) at 1/2 filling. The vertical and horizontal axes denote the valley and spin splittings respectively.}
\label{fig:ahc-tdbg}
\end{figure}
We first calculate the AHC at zero filling (filling 4 out of the 8 low-energy bands) and +1/2 filling (filling 6 out of the 8 low-energy bands) of  $AB$-$BA$ stacked TDBG at the twist angle $\theta\!=\!1.05^{\circ}$. In Fig.~\ref{fig:ahc-tdbg}(a) we show the the AHC of $AB$-$BA$ stacked TDBG at zero filling in the parameter space of $E_v$ and $E_s$. We see that the AHC is as large as $\sim -3.5\,e^2/h$ when the valley splitting $\sim\!3\,meV$, and gradually deceases with the increase of spin splitting. With the Hamiltonian Eq.~(\ref{eq:htdbg-eves}), the TDBG system is always metallic for $0\!\leq\!E_v\!\leq\!3\,$meV and 
$0\!\leq\!E_s\!\leq\!3\,$meV, thus the AHC is not quantized. In Fig.~\ref{fig:ahc-tdbg}(b) we show the AHC of $AB$-$BA$ stacked TDBG at +1/2 filling. Again, our calculations show that a small valley splitting $\sim\!3\,$meV would generate a substantial AHE with the AHC $\sim -3.4 e^2/h$.  Certainly the specific value of the AHC is sensitive to the details of the system. However, our calculations indicate that small valley splittings in $AB$-$BA$ stacked TDBG would lead to giant AHE, and such a feature of TDBG should be qualitatively correct. This is because such a property is a direct consequence of the valley Chern numbers and orbital ferroamgnetism of the low-energy bands in TDBG \cite{jpliu-tmg-arxiv19}, and it cannot occur in conventional spin ferromagnetic metals and insulators in which the AHE is generated through SOC as a perturbative effect. In the latter the AHC is proportional to the strength of SOC, which is typically much smaller than the bandwidth. As a result, the AHC is typically much smaller in magnitude than the diagonal conductivities. 
\begin{figure}
\includegraphics[width=3.5in]{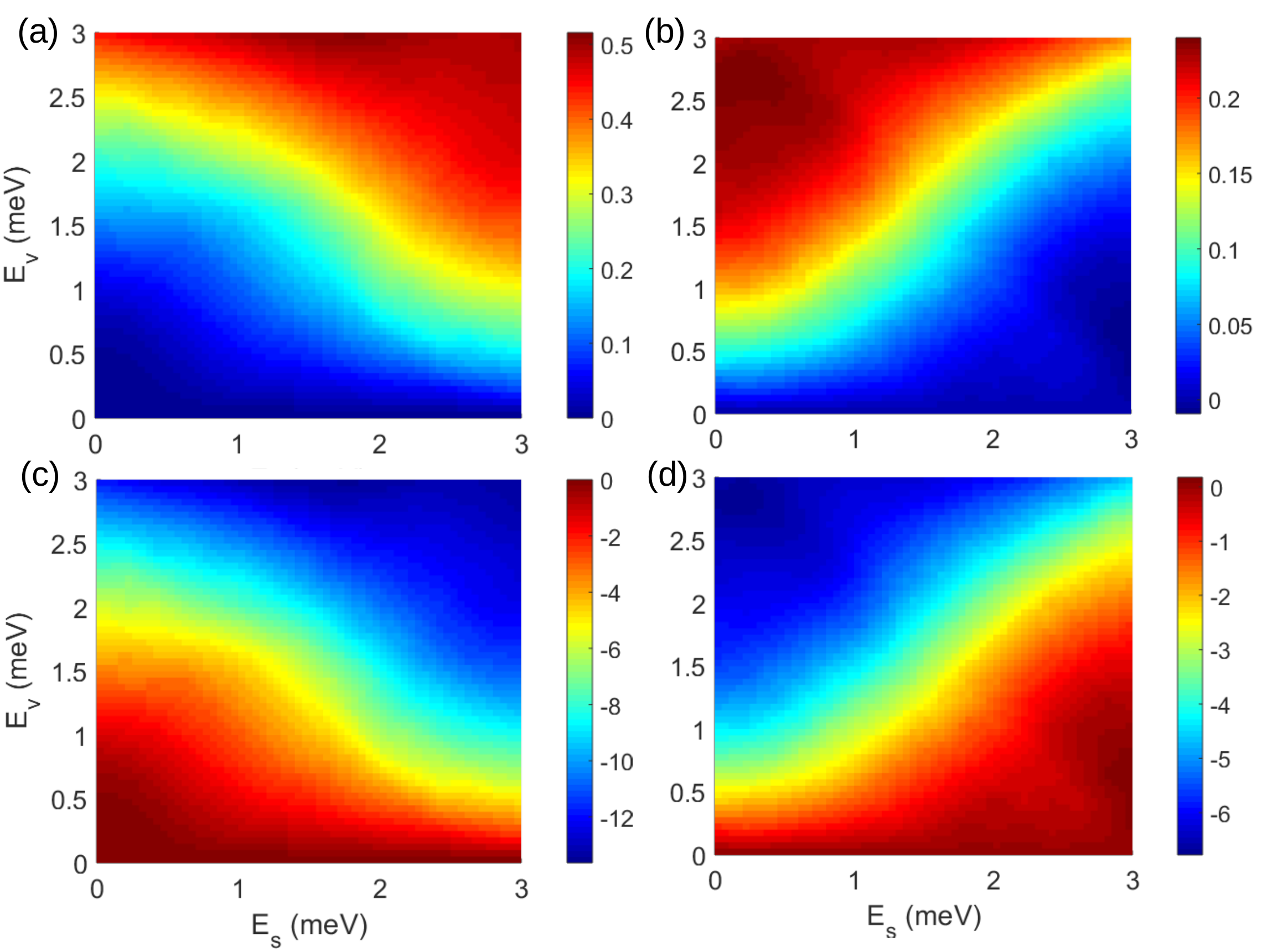}
\caption{Faraday and Kerr rotations of $AB$-$BA$ stacked TDBG: (a) Faraday angle at zero filling, (b) Faraday angle at 1/2 filling, (c) Kerr angle at zero filling, and (d) Kerr angle at 1/2 filling}
\label{fig:kerr-tdbg}
\end{figure}

The above argument also applies to magneto-optical phenomena. In Fig.~\ref{fig:kerr-tdbg}(a) and (b) we plot the Faraday rotations of $AB$-$BA$ stacked TDBG at $\theta\!=\!1.05^{\circ}$ at zero filling and $+1/2$ filling respectively, with the incident light frequency $\hbar\omega\!=\!0.05\,$eV. For zero filling (Fig.~\ref{fig:kerr-tdbg}(a)), the Faraday angle $\theta_F$ is largest around the upper right corner when $E_s\!\approx\!E_v\!\sim 2-3\,$meV, with the maximal $\theta_F\!\sim\!0.5^{\circ}$. For +1/2 filling (Fig.~\ref{fig:kerr-tdbg}(b)), $\theta_F$ is largest when $E_v\!\approx\!3\,$meV, and $E_s\!\sim\!0$\,meV, and the maximal $\theta_F\!\sim\!0.2^{\circ}$. By virtue of the orbital magnetism and nontrivial valley Chern numbers, small valley splittings $\sim\!2-3\,$meV would be strong enough to generate giant Faraday rotations.  In Fig.~\ref{fig:kerr-tdbg}(c) and (d) we plot the Kerr rotations of $AB$-$BA$ stacked TDBG for zero filling and $+1/2$ filling respectively. We find that the valley splittings and orbital magnetization in $AB$-$BA$ stacked TDBG would generate giant Kerr rotations $\sim -10^{\circ}$ at zero filling and $\sim -5^{\circ}$ at +1/2 filling, with the incident light frequency $\hbar\omega=0.05\,$eV.

\subsubsection{$AB$-$AB$ stacking}

On the other hand, in $AB$-$AB$ stacked TDBG, the orbital magnetization for each valley vanishes as a result of  $C_{2x}$ symmetry \cite{jpliu-tmg-arxiv19}, which implies that the AHC and the magneto-optical Kerr/Faraday rotations would vanish as well, as both effects are induced by the orbital magnetization. However, the $C_{2x}$ symmetry for each valley would be broken due to the presence of vertical electric fields,  which  gives rise to isolated topological flat bands with tuable valley Chern numbers \cite{ashvin-double-bilayer-arxiv19, koshino-tdbg-prb19, jpliu-tmg-arxiv19}. The non-zero valley Chern numbers of the topological flat bands are associated with valley contrasting orbital magnetizations, which may lead to substantial AHE and Kerr/Faraday rotations if the valley symmetry is broken.

Ferromagnetic insulating states have been observed in experiments at 1/2 filling of the isolated topological flat bands in $AB$-$AB$ stacked TDBG \cite{cao-double-bilayer-arxiv19, kim-double-bilayer-arxiv19, zhang-double-bilayer-arxiv19}. Such ferromagnetic insulating states have been proposed to be spin ferromagnetic states \cite{kim-double-bilayer-arxiv19, ashvin-double-bilayer-arxiv19}, which is not expected to exhibit any AHE nor magneto-optical effects due to the negligible SOC in graphene. However, at 1/4 or 3/4 filling of the isolated topological flat bands, it is possible to achieve a QAH state with both valley and spin polarizations. If such a state could be realized, then the system is expected to have significant Faraday and Kerr effects as in the cases of $AB$-$BA$ stacked TDBG and hBN-aligned TBG.


%
\subsection{Nonlinear optical properties}

\subsubsection{$AB$-$BA$ stacking}
\begin{figure}
\includegraphics[width=3.5in]{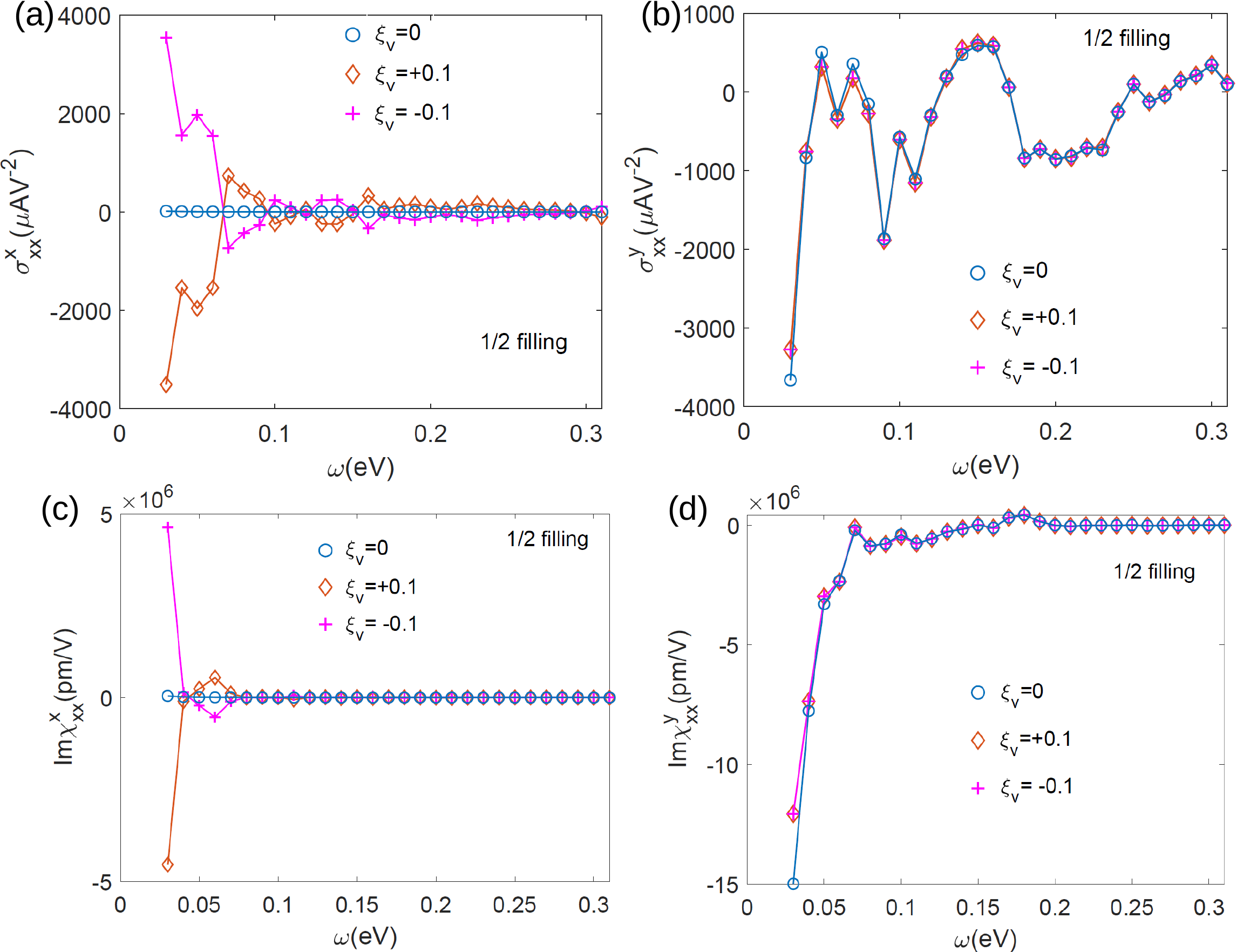}
\caption{(a)-(b), shift-current photoconductivities   of  $AB$-$BA$ stacked TDBG at 1/2 filling: (a) $\sigma^{x}_{xx}(0)$, and (b) $\sigma^{y}_{xx}(0)$. 
(c)-(d), the imaginary parts of  the SHG susceptibilities of $AB$-$BA$ stacked TDBG at 1/2 filling: (c) Im$\chi^{x}_{xx}(2\omega)$, and (d) Im$\chi^{y}_{xx}(2\omega)$. The blue circles, red diamonds, and magenta plus signs denote the cases of 0\%, +10\%  and -10\% valley polarizations respectively.}
\label{fig:shg-shift-tdbg}
\end{figure}
We continue to study the nonlinear optical properties of TDBG. Again, before going into the details, we first make symmetry analysis on the nonlinear photo-conductivity tensor. $AB$-$BA$ stacked TDBG has both $C_{2y}$ and $C_{3z}$ symmetries \cite{koshino-tdbg-prb19}. The $K$ and $K'$ valleys are invariant under $C_{3z}$ operation, but are interchanged with each other under $C_{2y}$ operation. Therefore, for each valley of $AB$-$BA$ stacked TDBG, there is only $C_{3z}$ symmetry, and the symmetry allowed form of the second-order photoconductivity tensor for each valley is already given by Eq.~(\ref{eq:shift-c3z}). 
However, in $AB$-$BA$ TDBG, the $C_{2y}$ symmetry would further enforce the photo-conductivities of  the $K$ and $K'$ valleys to the following form,
\begin{align}
&\sigma^{x}_{xx,0}(K)=-\sigma^{x}_{xx,0}(K')\;\nn
&\sigma^{x}_{xx,z}(K)=\sigma^{x}_{xx,z}(K')\;\nn
&\sigma^{y}_{xx,0}(K)=\sigma^{y}_{xx,0}(K')\;\nn
&\sigma^{y}_{xx,z}(K)=-\sigma^{y}_{xx,z}(K')\;,
\label{eq:shift-ABBA}
\end{align}
where $\sigma^{x}_{xx,0}$, $\sigma^{x}_{xx,z}$, $\sigma^{y}_{xx,0}$, $\sigma^{y}_{xx,z}$ are defined in Eq.~(\ref{eq:shift-general}).  
It follows  that if the two valleys remain degenerate,  $\sigma^{x}_{xx}=\sigma^{x}_{xx,0}+\sigma^{x}_{xx,z}M_z$ must vanish, because the  contributions from the opposite valleys (with opposite $M_z$) would exactly cancel each other. If the valley symmetry is broken,  the net orbital magnetization $M_z$ would be non-vanishing, leading to nonzero $\sigma^{x}_{xx}\sim \sigma^{x}_{xx,z} M_z$.  Thus the $\sigma^{x}_{xx}$ component of the nonlinear photoconductivity tensor  can be used as a probe to detect the  valley symmetry breaking and the associated orbital magnetization in $AB$-$BA$ stacked TDBG. Once the orbital $\mathcal{T}$ symmetry (valley symmetry) is spontaneously broken, the $\sigma^{x}_{xx}$ component would be linearly proportional to the orbital magnetization of the system, which is expected to exhibit hysteresis behavior in response (only) to the out-of-plane magnetic field.

In Fig.~(\ref{fig:shg-shift-tdbg})(a) we plot the shift-current photoconductivity $\sigma^{x}_{xx}(0)$ at  +1/2 filling. The blue circles, red diamonds, and magenta plus signs represent the situations with the valley polarization $\xi_v=0$, $\xi_v=0.1$, and $\xi_v=-0.1$ respectively. Clearly, when the valley symmetry is preserved (valley polarization $\xi_v=0$), $\sigma^{x}_{xx}$ identically vanishes at any frequency, as the contributions from the two valleys exactly cancel each other. On the other hand, with $\pm10\%$ of valley polarizations ($\xi_v\!=\!\pm 0.1$), $\sigma^{x}_{xx}$ can be as large as $\pm 10^{3}\,\mu$AV$^{-2}$ at relatively low frequencies $\hbar\omega\lessapprox 0.1\,$eV  as shown in Fig.~(\ref{fig:shg-shift-tdbg})(a). Moreover, $\sigma^{x}_{xx}$ are opposite for opposite valley polarizations, because $\sigma^{x}_{xx}$ is proportional to the total orbital magnetization as argued above. At higher frequencies $\hbar\omega\gtrapprox 0.1\,$eV, $\sigma^{x}_{xx}$ gradually decreases to $\sim 10^{2}\mu\,$A V$^{-2}$. In Fig.~(\ref{fig:shg-shift-tdbg})(b) we show the shift-current photoconductivity $\sigma^{y}_{xx}(\omega=0)$ at 1/2 fillings with the valley polarizations $\xi_v=0$, $+0.1$ and $-0.1$. Clearly the valley polarization does not significantly  change the value of $\sigma^{y}_{xx}$, which is always on the order of $10^{3}\,\mu$A V$^{-2}$ for $\hbar\omega\lessapprox 0.1\,$eV, and decrease to  $\sim 10^{2}\,\mu$A V$^{-2}$ at higher frequencies. It is worthwhile to note that $\sigma^{y}_{xx}$ are nearly identical for $\xi_v\!=\!\pm0.1$, which is expected according to Eq.~(\ref{eq:shift-ABBA}).

In Fig.~\ref{fig:shg-shift-tdbg} (c)-(d) we plot the imaginary part of the SHG susceptibilities $\chi^{x}_{xx}(2\omega)$ and $\chi^{y}_{xx}(2\omega)$ at 1/2 filling of $AB$-$BA$ stacked TDBG. Again, the blue circles, red diamonds, and magenta plus signs denote the cases with 0\%, +10\% and -10\% valley polarizations respectively. We see that $\textrm{Im}\chi^{x}_{xx}(2\omega)$ is giant ($\sim \pm 10^6\,$pm/V) when $\xi_v=\pm 0.1$ at low frequencies, but vanishes when $\xi_v=0$. 
Again, since $\chi^{x}_{xx}(2\omega)$ is directly proportional to the total orbital magnetization of the system, it has opposite signs for opposite valley polarizations.
To the contrary, $\textrm{Im}\chi^{y}_{xx}(2\omega)$ seems to be not sensitive to the valley polarizations, which is  always on the order of $10^6\,$pm/V for $\hbar\omega\lessapprox\,0.1$eV, and gradually decreases to  $10^3-10^4\,$pm/V at higher frequencies.

\subsubsection{$AB$-$AB$ stacking}

In $AB$-$AB$ stacked TDBG, there are $C_{3z}$ and $C_{2x}$ symmetries for each valley.  As a result, the only non-vanishing photoconductivity components for each valley are $\sigma^{x}_{xx,0}$ and $\sigma^{y}_{xx,z}$ as defined in Eq.~(\ref{eq:shift-general}), i.e.,
\begin{align}
&\sigma^{x}_{xx,0}(K^{\mu})=-\sigma^{x}_{yy,0}(K^{\mu})=-\sigma^{y}_{xy,0}(K^{\mu})=-\sigma^{y}_{yx,0}(K^{\mu})\;\nn
&\sigma^{y}_{xx,z}(K^{\mu})=\sigma^{x}_{xy,z}(K^{\mu})=\sigma^{x}_{yx,z}(K^{\mu})=-\sigma^{y}_{yy,z}(K^{\mu})\;,
\end{align}
where $\mu=\pm 1$ refers to the valley index, with $K^{-}=K$, and $K^{+}=K'$.
However, the $C_{2x}$ symmetry further enforces that $M_z\!=\!0$ for each valley, which implies $\sigma^{y}_{xx}=\sigma^{y}_{xx,z}\,M_z$ must vanish. The $C_{2x}$ symmetry can be broken by vertical electric fields, which would allow for non-vanishing but valley-contrasting orbital magnetizations. If the valley symmetry is further broken either spontaneously by Coulomb interactions or by external magnetic fields, then the orbital magnetization would contribute to $\sigma^{y}_{xx}$ in such a way that $\sigma^{y}_{xx}$ would exhibit hysteresis behavior under out-of-plane magnetic fields. To be specific, in the presence of vertical electric fields, for the valley $K^{\mu}$, $\sigma^{y}_{xx}(K^{\mu})$ is expressed as
\begin{align}
\sigma^{y}_{xx}(K^{\mu})=\sigma^{y}_{xx,0}(K^{\mu})+\sigma^{y}_{xx,z}(K^{\mu})\, M_{z}(K^{\mu})\;,
\label{eq:sigmayxx}
\end{align}
Note that the coefficients $\sigma^{y}_{xx,0}(K^{\mu})$ and $\sigma^{y}_{xx,z}(K^{\mu})$ do not change signs under time-reversal operation, and they are dependent on the valley indices only if the bandstructures and/or chemical potentials of the two valleys are different. 
%
%
Then we consider the situation that the system has nonzero valley splittings $\pm E_v$ (see Eq.~(\ref{eq:htdbg-eves}).  When the valley splitting is $+E_v$, the orbital magnetization of the $K$ and $K'$ valleys are denoted as $M_z^{-}$ and $M_{z}^{+}$ respectively; if the valley splitting is reversed to $-E_v$, then the orbital magnetizations of the $K$ and $K'$ valleys would become $-M_{z}^{+}$ and $-M_z^{-}$. Plugging these relationships into Eq.~(\ref{eq:sigmayxx}), one obtains
\begin{align}
&\sigma^{y}_{xx}(+E_v)=\sum_{\mu=\pm 1}\sigma^{y}_{xx,0}(\mu E_v)+\sum_{\mu=\pm1}\sigma^{y}_{xx,z}(\mu E_v) \,M_z^{\mu}\;\nn
&\sigma^{y}_{xx}(-E_v)=\sum_{\mu=\pm 1}\sigma^{y}_{xx,0}(-\mu E_v)-\sum_{\mu=\pm 1}\sigma^{y}_{xx,z}(-\mu E_v)\, M_z^{-\mu}\;,
\label{eq:syxx-total}
\end{align}
where $\sigma^{y}_{xx}(\pm E_v)$ stands for the total $\sigma^{y}_{xx}$ with $\pm$ valley splittings, summing over the contributions from the two valleys. Subtracting $\sigma^{y}_{xx}(+Ev)$ by $\sigma^{y}_{xx}(-E_v)$ would eliminate the $\sigma^{y}_{xx,0}$ term, leaving the term that is proportional to the total orbital magnetization of the system, i.e., 
\begin{align}
&\Big(\,\sigma^{y}_{xx}(+E_v)-\sigma^{y}_{xx}(-E_v)\,\Big)/2\;\nn
=&\sigma^{y}_{xx}(-E_v)M_z^{-}+\sigma^{y}_{xx}(+E_v)M_z^{+}\;\nn
\approx &\sigma^{y}_{xx}(0)\,(M_z^{-}+M_z^{+})\;,
\label{eq:syxx-orbital}
\end{align}
where $M_z^{+}+M_z^{-}$ is the total orbital magnetization of the system with $+E_v$ valley splitting. In the last line of Eq.~(\ref{eq:syxx-orbital}) we have assumed $\sigma^{y}_{xx}(-E_v)\approx\sigma^{y}_{xx}(0)\approx\sigma^{y}_{xx}(E_v)$ for small valley splittings. Similar argument also applies to $\sigma^{x}_{xx}$. When $C_{2x}$ symmetry is broken by vertical electric fields, the $\sigma^{x}_{xx,z}$ parameter would be non-vanishing for each valley, such that the orbital magnetization would also contribute to $\sigma^{x}_{xx}$. One can also extract the term proportional to $M_z$ by subtracting $\sigma^{x}_{xx}(-E_v)$ from  $\sigma^{x}_{xx}(+E_v)$.

\begin{figure}
\includegraphics[width=3.5in]{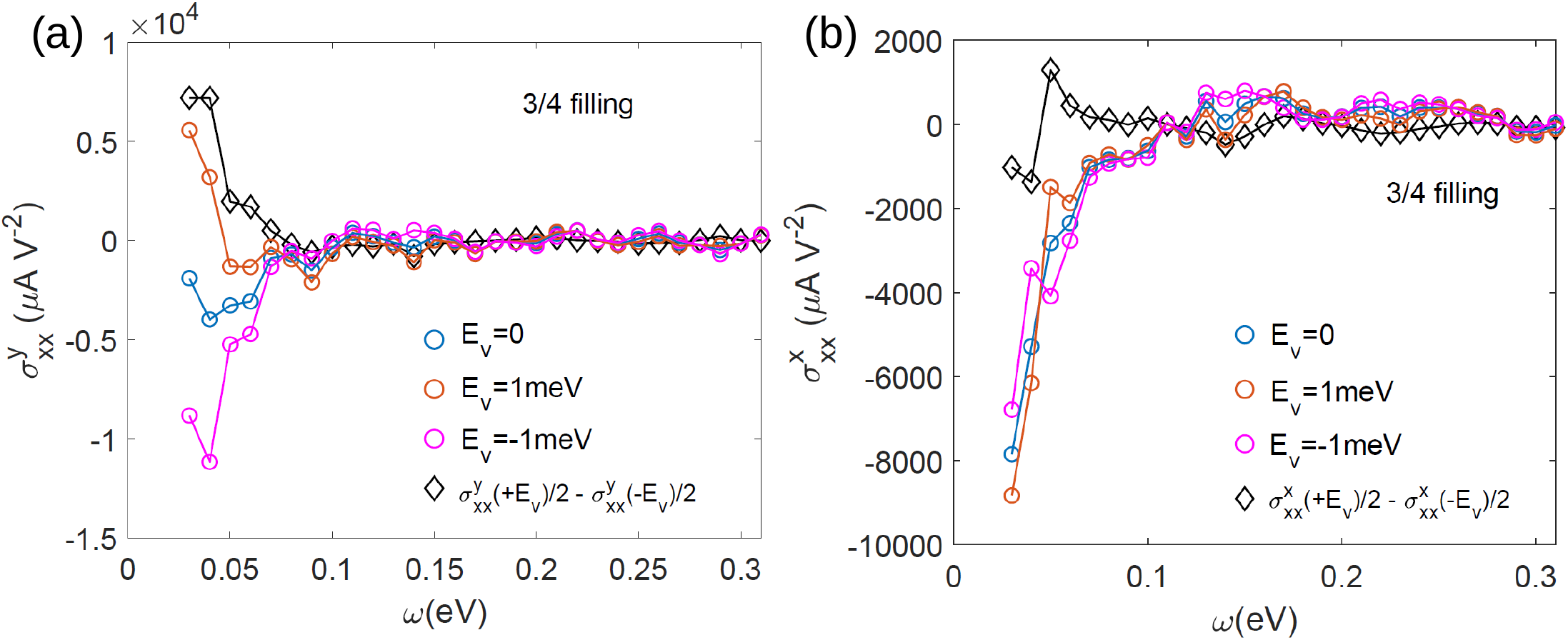}
\caption{(a) The shift-current photoconductivity $\sigma^{y}_{xx}$, and (b) $\sigma^{x}_{xx}$, for 3/4 filled $AB$-$AB$ stacked TDBG with vertical electric fields at the twist angle $\theta\!=\!1.05^{\circ}$. The blue, red, and magenta circles represent the situations with the valley splitting $E_v\!=\!0$, $+1\,$meV, and $-1\,$meV respectively. The black diamonds represent $\sigma^{y}_{xx}(+E_v)/2-\sigma^{y}_{xx}(-E_v)/2$ in (a), and $\sigma^{x}_{xx}(+E_v)/2-\sigma^{x}_{xx}(-E_v)/2$ in (b), with $E_v\!=\!1\,$meV. }
\label{fig:shift-shg-abab}
\end{figure}

In Fig.~\ref{fig:shift-shg-abab} we show the shift-current response of $AB$-$AB$ stacked TDBG with vertical electrostatic potential drop  $U_d=0.045\,$eV across the four layers. In Fig.~\ref{fig:shift-shg-abab}(a) we present the shift-current photoconductivity $\sigma^{y}_{xx}(0)$ at 3/4 filling of the isolated conduction flat band at $\theta=1.05^{\circ}$.  The blue, red and magenta circles represent the cases with valley splitting $E_v\!=\!0$, $+1\,$meV and $-1\,$meV respectively. The black diamonds denote $\sigma^{y}_{xx}(+E_v)/2-\sigma^{y}_{xx}(-E_v)/2$  with $E_v=1\,$meV, which extracts the orbital-magnetization contribution to $\sigma^{y}_{xx}$ as explained in Eqs.~(\ref{eq:sigmayxx})-(\ref{eq:syxx-orbital}). We see that $\sigma^{y}_{xx}$ is actually dominated by the $\sigma^{y}_{xx,z} M_z$ term, which is expected to show remarkable hysteresis loops under out-of-plane magnetic fields. In Fig.~\ref{fig:shift-shg-abab}(b) we plot the dependence of $\sigma^{x}_{xx}(0)$ on the light frequency $\omega$ at the same filling and the same twist angle. Clearly $\sigma^{x}_{xx}(0)$ is not changed too much by the valley splitting $E_v$, indicating that the orbital-magnetization contribution ($\sigma^{x}_{xx,z} M_z$) plays a minor role in  $\sigma^{x}_{xx}(0)$.




\section{Conclusion}

To summarize,  we  have systematically studied the  anomalous Hall effect, magneto-optical properties, and nonlinear optical properties of  hBN-aligned TBG and TDBG. We have studied the dependence of AHC on the valley and spin splittings in hBN-aligned TBG, and found that the AHE can be engineered using in-plane magnetic fields.
In additional to AHE, we also show that there exists giant magneto-optical effect by virtue of the valley-symmetry breaking and orbital magnetizations. We propose that the Faraday and Kerr rotations may be a powerful tool to detect the presence of orbital magnetism in twisted graphene systems. Moreover, we have also studied the nonlinear optical responses, i.e., the shift current and second harmonic generation  in both hBN-aligned TBG and TDBG. Our calculations indicate that both systems exhibit colossal nonlinear optical responses.  To be specific, the shift-current photoconductivity $\sigma^{c}_{ab}(0)$ is on the order of $10^{3}\,\mu$A/V$^2$,  and the SHG susceptibility $\chi^{c}_{ab}(2\omega)$ is on the order of $10^6$\,pm/V in the terahertz frequency regime. Such gigantic nonlinear optic responses are by virtue of the inversion symmetry breaking, the presence of the low-energy flat bands, and  the small excitation gaps in the twisted graphene systems.  In TDBG with $AB$-$BA$ stacking, we propose that the non-vanishing valley polarization and orbital magnetization ($M_z$) are associated with $C_{2y}$ crystalline symmetry breaking, which would generate  a new component of the nonlinear photoconductivity $\sigma^{x}_{xx}\sim M_z$; while in $AB$-$AB$ stacked TDBG with vertical electric fields, the valley polarization and orbital magnetization would make significant contributions to $\sigma^{y}_{xx}$. 
  These new components of photoconductivities generated by the orbital magnetizations would exhibit notable hysteresis behavior in response to out-of-plane magnetic fields, and may be considered as strong and robust experimental evidence for the valley polarized state and orbital magnetism in the TDBG system. Our work is a significant step forward in understanding the optical properties of the twisted graphene systems, and may provide useful guidelines for future experimental works.

\acknowledgements
J.L. and  X.D. acknowledge  financial support from the Hong Kong Research Grants Council (Project No. GRF16300918). We thank Shiwei Wu, Hongming Weng, Yang Zhang,  Hiroaki Ishizuka, and Liuyan Zhao for invaluable discussions.

\bibliography{tmg}
\end{document}